\documentclass[aps,preprint,onecolumn,eqsecnum]{revtex4}
\usepackage{dcolumn}
\usepackage{bm}
\usepackage{amssymb}
\usepackage{amsmath}
\usepackage{amsfonts}

\usepackage[]{graphicx}
\begin{document}
\bibliographystyle{prsty}
\title{About Wigner Friend's and Hardy's paradox in a  Bohmian approach: a comment of  `Quantum theory cannot' consistently describe the use of itself'}
\author{ Aur\'elien Drezet $^{1}$}
\address{(1) Institut NEEL, CNRS and Universit\'e Grenoble Alpes, F-38000 Grenoble, France 
}
\begin{abstract}
This is an analysis of  the recently published article `Quantum theory cannot consistently describe the use of itself' by D. Frauchiger and R. Renner~\cite{1}. Here I decipher the paradox and analyze it from the point of view of de Broglie-Bohm hidden variable theory (i.e., Bohmian mechanics). I also analyze the problem from the perspective obtained by the Copenhagen interpretation (i.e., the Bohrian interpretation) and show that both views are self consistent and do not lead to any contradiction with a `single-world' description of quantum theory.
\end{abstract}

\maketitle
\section{Hardy's paradox}
 \label{sec:1}
\indent The claim of this article is that the recently published article \cite{1,1b} by D. Frauchiger and R. Renner  about   Wigner's Friends~\cite{0} and entanglement is mainly a rephrasing of the beautiful Hardy paradox~\cite{3,3a} about quantum non-locality without inequality (for a clear and nice derivation see also \cite{3b} by S. Goldstein; see also the Greenberger-Horne-Zeilinger (GHZ) paradox~ \cite{GHZ}). The authors of \cite{1} recognize the importance of Hardy's letter in their own analysis but the argument is written in such a way that the relation is no immediately transparent. My aim is here to clarify this point from the point of view of Bohmian mechanics (i.e. de Broglie Bohm interpretation). During the analysis I will also consider the perspective taken by various Bohrians (i.e. adepts of the Copenhagen interpretation) and show the intimate relation this has with a Bohmian perspective. The problem is also connected to a work by C. Brukner~\cite{2,2b} where an argument similar to~\cite{1} is obtained but based on  Bell's inequality for the singlet spin state (and for the GHZ paradox~\cite{2,GHZ}). Our analysis also apply to this work.\\
 \indent The argument of \cite{1,3} is the following: Take two Qbits and entangle them in such a way  as to obtain the pure quantum state:
\begin{eqnarray}
|\Psi\rangle=\frac{1}{\sqrt{3}}|h,\downarrow\rangle +\frac{1}{\sqrt{3}}|t,\downarrow\rangle+\frac{1}{\sqrt{3}}|t,\uparrow\rangle,
\label{eq1}
\end{eqnarray}
where $h$ and $t$ refer to  head and tail of the quantum coin used in \cite{1} and $\downarrow,\uparrow$ to the spin state of the system S consider in the same article. Hardy's paradox comes when we consider 4 kinds of measurements which 
are in counter factual conflicts with each other and based on the nonlocality and contextuality of quantum lead to a contradiction or paradox.\\
\indent Consider indeed  quantum states~\cite{comment} $|\overline{ok}\rangle=\frac{|h\rangle-|t\rangle}{\sqrt{2}}$, $|\overline{fail}\rangle=\frac{|h\rangle+|t\rangle}{\sqrt{2}}$ (basis $\overline{W}$), $|ok\rangle=\frac{|\uparrow\rangle-|\downarrow\rangle}{\sqrt{2}}$, $|fail\rangle=\frac{|\uparrow\rangle+|\downarrow\rangle}{\sqrt{2}}$  (basis W) used in \cite{1}. We s start  with the experiment where we measure if the first Qbit is in the state $h$ or  $t$  and  use basis $W$ for measuring  the second Qbit.\\
\indent From Eq.~\ref{eq1} if we get $|t\rangle$ then we must have $|fail\rangle$.  However if we measure $|h\rangle$ we can get with the same probability $|ok\rangle$ or $|fail\rangle$~\cite{5}.    Therefore, we have the logical  inference:\\
`\textit{if we measure $|ok\rangle$ for the second Qbit then we have  $|h\rangle$ for the first one}' (we call this inference $\textbf{I}$).\\ 
For symmetric reasons \cite{5}  in a different experiment by using the basis $\overline{W}$  and $\{|\uparrow\rangle,|\downarrow\rangle\}$ we get the inference:\\
 `\textit{if we measure $|\overline{ok}\rangle$ for the second Qbit then we have  $|\uparrow\rangle$ for the first one}' (we call this inference $\overline{\textbf{I}}$).\\
 Apriori, in a local world, if the two Qbits are far way we should  from $\textbf{I}$ and $\overline{\textbf{I}}$ deduce:\\
 ` \textit{if we could observe  the Qbits in the state $|ok,\overline{ok}\rangle$ then we should also have in a counterfactual reasoning a contribution $|h,\uparrow\rangle $ in the initial state}'.\\
  However, the state of Eq.~\ref{eq1} doesn't contain any contribution $|h,\uparrow\rangle $ therefore  from this apriori correct reasoning we should never observe a state like $|ok,\overline{ok}\rangle$ in our measurement (we call this inference $\textbf{I}+\overline{\textbf{I}}$). An equivalent way to obtain this result is to say that we have the chain of logical implications\cite{2}: $\overline{ok}\rightarrow \uparrow\rightarrow t\rightarrow fail$.  However, and this is Hardy's paradox, if you actually \textit{do} the measurement in the bases $W$ and $\overline{W}$ then we will get with a probability of occurrence  
\begin{eqnarray}
P_\Psi(ok,\overline{ok})=\frac{1}{12}\label{eq2}
\end{eqnarray} the state  $|ok,\overline{ok}\rangle$ which is contradicting the previous counterfactual reasoning $\textbf{I}+\overline{\textbf{I}}$. This is a remarkable proof of 'nonlocality' without inequality showing that  if we want to give a mechanical explanation of quantum mechanics (i.e., with hidden variables) then we we must include  a nonlocal action at a distance which prohibits us to take too seriously the previous inference mixing $\textbf{I}$ and $\overline{\textbf{I}}$. In a Bohmian approach for example~\cite{8a,8}, there is an additional quantum force or potential acting nonlocally on the two Qbits when we use the measurement bases $W$ and $\overline{W}$. This  additional quantum force is context dependent meaning that the experiments leading to inferences $\textbf{I}$, $\overline{\textbf{I}}$ and  to  Eq.~\ref{eq2} are not possible in the same context and imply different hidden variable dynamics and quantum forces. The contradiction results (in the Bohmian approach) from forgetting the nonlocal and contextual quantum force acting on particle trajectories.\\
\section{Wigner's friends and Hardy's paradox }
 \label{sec:2}
\indent So what is new in Ref.~\cite{1}? The authors actually introduces four agents or observers and develop a story plot  similar in philosophy to the famous Wigner friend paradox~\cite{0} but now based on the nonlocality a la Hardy and involving two Wigner friends. In other words, they introduce macroscopic devices with memories (using  John Bell unconventional convention let call them PhD students). Two first agents $F$ and $\overline{F}$ are supposed to be strongly entangled with the Qbits and somehow measure the states of the two Qbits in the basis $\{|\uparrow\rangle,|\downarrow\rangle\}$ for $F$  and $\{|h\rangle,|t\rangle\}$ for $\overline{F}$. Now, in the story plot the quantum coin and  $\overline{F}$ are in a Lab $\overline{L}$ isolated from the rest of the Universe up to small communicating channels.  We also  suppose the same  for the quantum spin  and the observer F which are together in a lab  $L$ also well isolated from the Universe and in particular from $\overline{L}$. The quantum state Eq.~\ref{eq1} is thus becoming a statement about the entanglement of the two labs. For example the basis vector $|h,\downarrow\rangle$  means  that  the observer  $F$ and its local environment in $L$ is in the state $\downarrow$ whereas the observer $\overline{F}$ in his lab  $\overline{L}$ is in the state $h$ (see Eq.~2 of \cite{1}).\\
\indent Now, Eq.~\ref{eq1} means that from Quantum theory we preserve phase coherence between the various alternatives or branches  which are $(h,\downarrow)$, $(t,\downarrow)$ and $(t,\uparrow)$. In a world where there is a kind of Heisenberg cut between the quantum-uncollapsed Universe and the classical looking like collapsed Universe these 3 alternatives can not see each other and we could replace every thing by density matrices. However, in \cite{1} the entanglement is supposed to be preserved (no decoherence occured yet), quantum mechanics is supposed to be universally valid (there is no objective collapse), and some super observers called Wigner friends  $W$ and $\overline{W}$ are recording the  quantum states of labs  $L$ and $\overline{L}$.\\ 
\indent In this new level of description observers $W$ and $\overline{W}$ can communicate meaning that the world is classical or collapsed (i.e. decohered) for them. Still, they are able to make projective measurements on the basis vectors $ok$, $fail$ and $\overline{ok}$, $\overline{fail}$ which are macroscopic superposition of observers $F$, and $\overline{F}$ quantum states. Alternatively, $W$ and $\overline{W}$ could record the states of $F$, and $\overline{F}$ in the $h,t$ and $\uparrow,\downarrow$ bases. Now, all these experiments are defining contexts which are sometimes incompatibles and  we again  have the complete Hardy paradox with the contradiction surrounding statement $\textbf{I}+\overline{\textbf{I}}$ and Eq.~\ref{eq2}.  Every thing is the same but now every thing is macroscopic and therefore looks even more fantastic. Inferences $\textbf{I}$: `$ok\rightarrow h$' and $\overline{\textbf{I}}$: `$\overline{ok}\rightarrow \uparrow$' now have a macroscopic meaning involving PhD students  
and the conclusion  $\textbf{I}+\overline{\textbf{I}}$: `$ok,\overline{ok}\rightarrow h,\uparrow$' looks also natural from a classical perspective. But of course we are no classical here  if we preserve the phases and  if Wigner friends can use $F$ and $\overline{F}$ as if they were simple Qbits. Therefore,  we have no reason to believe in $\textbf{I}+\overline{\textbf{I}}$. For the super observers $W$ and $\overline{W}$  the paradox thus dissolves since  nonlocality precludes the use of a common experimental context for $\textbf{I}$, $\overline{\textbf{I}}$ and Eq.~\ref{eq2} in agreement with \cite{3}.\\ \indent Nevertheless, agents $F$ and $\overline{F}$ would no be happy to be treated as simple Qbits:   for sure they would disagree with the idea that they are in a quantum superposition, i.e., entangled with each other. Don't forget all these paradoxes and issues with Schrodinger cats and Wigner Friends come from the strict application of Bohr interpretation adapted to an experimentalist in the lab and dealing with atoms or photons. For Bohr, a spin in a quantum superposition has no clear description before the observation (I speak about description no about ontological existence which is a different thing).   Therefore, for super observers $W$ and $\overline{W}$ the less-super observers $F$ and $\overline{F}$ with PhD families, (Schrodinger) cats etc.. are  in the previous story plot also in a `foggy' state (Wheeler used the term `Great Smoky Dragon' as a metaphor) while $F$ and $\overline{F}$ will no accept it. The problem is no new:  what is the meaning of me being in a quantum state happy and unhappy $|\ddot{\cup}\rangle+|\ddot{\cap}\rangle$ ? How do I feel in  such a state?  
What happens to a conscious cat~\cite{7} in a quantum interferometer?  In my humble opinion the only self-consistent interpretation actually available for answering all these questions (i.e. ontological and epistemic) is the Bohmian one~\cite{7}. Only in this mathematically sharp interpretation can we write quantum superpositions and still have a clear ontological description  of both microscopic and macroscopic systems even without observer (this doesn't however preclude the development of a better theory or of an empirically equivalent one). A quantum cat can  be in a interferometer and still follow one single trajectory while a quantum guiding wave (or empty wave: Bohmians are still debating about it) goes through  many paths. The Many Worlds Interpretation is also a candidate for an ontological theory but in~\cite{7} we showed that this approach collapses completely because it can not seriously describe what is a probability.  In the Bohmian approach there is no problem into defining a chain of observers a la von Neumann or Wigner: each observer will be real and possess a univocal state of affair. Objective collapse theories such as spontaneous collapse a la  Pearle, GRW, Penrose, Diosi or Tumulka\cite{g}  (may be should we also include in this list Wigner's approach where the mind-body frontier plays a key role in the wave-function collapse) are also interesting  but they require a new level of unknown physics going beyond far quantum mechanics. In the example given in \cite{1} Bohmian mechanics predicts that the superposition Eq.~\ref{eq1} will involve (extremely weak) nonlocal quantum forces which could in the hypothetical scenario lead indeed to the rejection of $\textbf{I}+\overline{\textbf{I}}$ and to the empirical justification of Eq.~\ref{eq2}. Still $F$ and $\overline{F}$ will have  clearly defined trajectories without any foggy elements  (this is what Einstein called a complete theory). \\ 
\indent In order to clarify the role of Bohmian mechanics for understanding the paradox discussed in \cite{1} we should first go back to the structure of the axiomatic discussed in this work. 
\subsection{Q,C,S axioms in \cite{1}: How an agent should apply Quantum mechanics?}
 \label{sec:2a}
\begin{figure*}[hbtp]
\includegraphics[width=1\columnwidth]{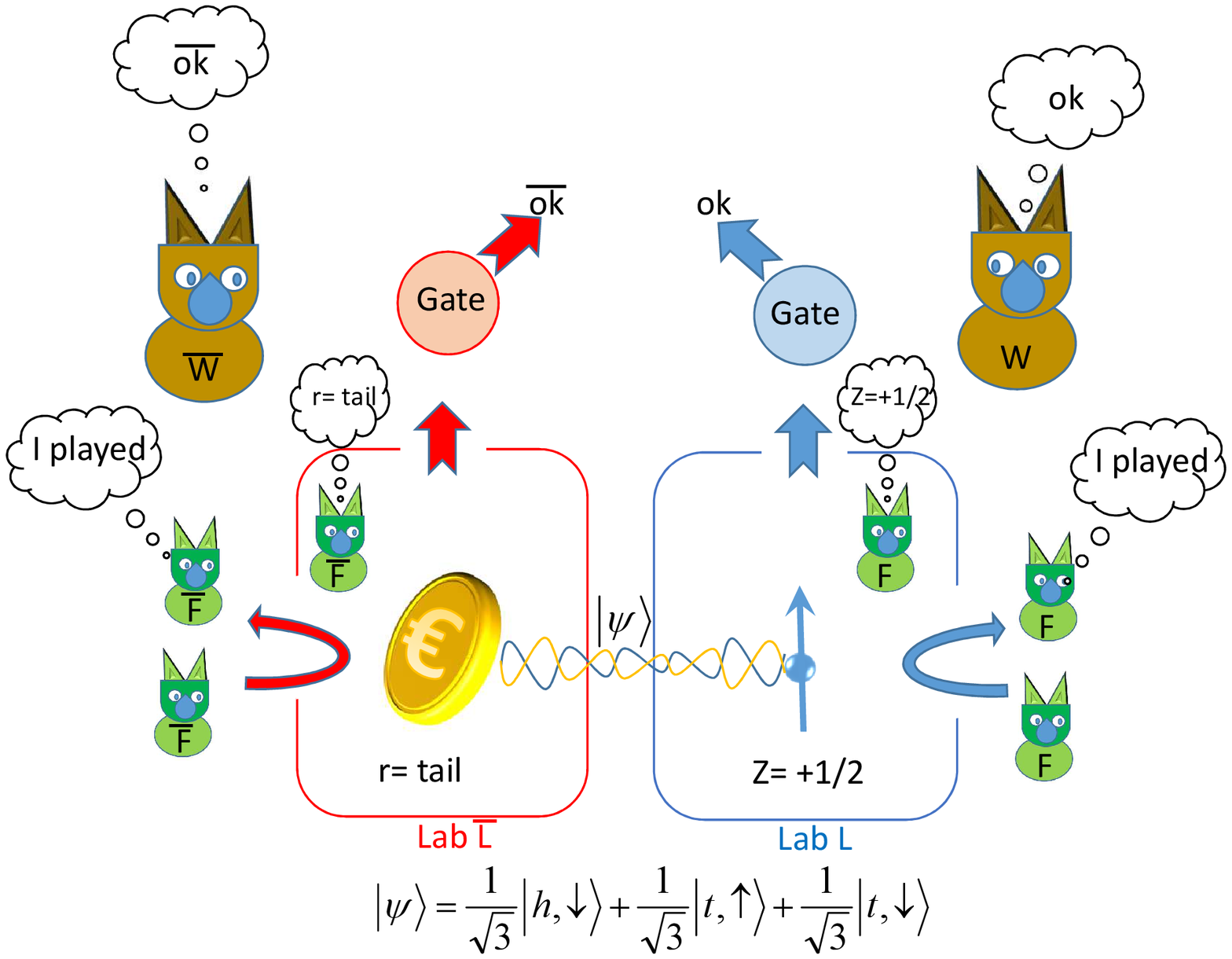}\\
\caption{ Sketch of the Gedanken experiment  proposed in \cite{1} but here  including a quantum memory device a la Deutsch~\cite{10}.  During a time $T$ the observers $F$ and $\overline{F}$ can interact with the two entangled Qbits in a Hardy state (see Eq.~\ref{eq1}). After this time they partially erase their quantum memory and only remember having obtained a definite outcome.  Subsequently, the Wigners $W$ and $\overline{W}$ can do different projective measurements by using linear gates or beam splitters on the two entangled Qbits as explained in sections \ref{sec:1} and \ref{sec:2b}. We emphasize that on the figure the choice for the outcomes obtained by the different observers are made in reference to panel (a) of Fig.~\ref{fig3} where Bohmian trajectories are used to graphically represent hidden-variable paths violating Hardy's non contextual axiom $\textbf{I}+\overline{\textbf{I}}$. }
\label{fig0}
\end{figure*}

 \cite{1} uses three assumptions (Q)-quantum mechanics is valid, (C)-mutual consistency between observers is required, and (S)-self consistency for an observer is also imposed  which  are leading to a no-go theorem. While the definition of these 3 assumptions is relatively clear the application of the rules to the problem discussed in \cite{1} is ambiguous.  The problem is with the definition of the observers $F$ and $\overline{F}$  and even more  with the use they do of quantum mechanics. As I wrote an observer is a macroscopic device with a memory (here a quantum memory). Of course an observer will not stay or live for ever and therefore  we must at least admit that during  the experiment the observer memory should not be quantum-erased. Quantum-erased should not be confused with simply  `destroyed'  because  in general destruction (by heat)  transfers some information to the environment which is enough to decohere the various quantum branches and thus to keep a track of the observation and generate a which path information. This criterion  is directly applicable to  $W$ and $\overline{W}$  but not to $F$ and $\overline{F}$ because the states $|\overline{fail}\rangle=\frac{|h\rangle+|t\rangle}{\sqrt{2}}$ etc...  are all linear combinations of the primary states $|h\rangle$, $|t\rangle$ etc.... If we dont want to ask our self what is meaning of  $|\ddot{\cup}\rangle+|\ddot{\cap}\rangle$ we can alternatively  imagine (see Fig.~\ref{fig0}) that the observers  $F$ and $\overline{F}$  before the experiment have an empty memory $|`\emptyset'\rangle_F$, $|`\emptyset'\rangle_{\overline{F}}$. In the first step of the experiment  they interact locally with their respective Qbits  and during a time $T$ can keep a memory of that result (in the bases $(h,t)$ and $(\downarrow,\uparrow)$) and can think and meditate about  the experiment (this is needed in the proposal \cite{1} where agents try to obtain conclusions about the results and outcomes of other agents).  After this time $T$  their memories are quantum-erased  and the observers $F$ and $\overline{F}$ leave the labs $L$ and $\overline{L}$ to discuss with $W$ and $\overline{W}$. For example, for observer $F$ watching the spin  we can have a branch: \begin{eqnarray}|`\emptyset'\rangle_{F}|\uparrow\rangle_S\rightarrow|`\uparrow'\rangle_{F}|\uparrow\rangle_S\rightarrow|`\emptyset'\rangle_F|\uparrow\rangle_S.\label{truc}\end{eqnarray} After this  operation the observers  $W$ and $\overline{W}$ can do their manipulation on the Qbits Eq.~\ref{eq1} in the bases $(ok, fail)$, $(\overline{ok}, \overline{fail})$ and we go back to the discussion of Hardy's experiment\cite{3}.\\
\indent D. Deutsch, introduced this idea in~\cite{10} where he discussed a similar paradox with the double slit experiment. Deutsch used an important novelty: the memory  doesn't need to be completely erazed.  Only the exact value of the outcome is erased  and the agent can keep in memory the fact that he or she got a single definite outcome. This already introduces a kind of paradox for those who believe that an observer must necessarily collapse the wave function. 
Importantly, for a Bohrian there is no problem and no paradox a la Deutsch~\cite{10} since the fact that $F$ and $\overline{F}$ lose their memory is saving the consistency of the principle of complementarity which prohibits you to have at the same time information associated with different conflicting experimental contexts.  Of course, $F$ and $\overline{F}$ will be offended to be not trusted by $W$ and $\overline{W}$. Indeed, they actually participated to the experiment and even if they don't remember which result or outcome they got  at least   they  could keep the memory  that they got \textit{a single} result~\cite{10,2} and thus this seems to go beyond what Bohrians claim. In \cite{1} the observers should have the ability to get and lose a (partial) quantum-memory and this is necessary if we need to exploit the phase relation and nonlocality contained in Eq.~\ref{eq1}. D. Deutsch claimed that this quantum memory is already in conflict with those who think that a macroscopic observer should necessarily collapse the wave function~\cite{10}. This corresponds to a specific but old reading of Bohrian or Wheelerian philosophy in which `no elementary quantum phenomenon is a phenomenon until it is brought to a close by an irreversible act of amplification' and where this irreversible act of amplification is possibly associated with the mind-body boundary of observers. \\ 
\indent Now, one of the issue in \cite{1} is that the observers $F$ and $\overline{F}$ make some deductions and inferences by applying (Q) and (C). These deductions can only be done during the period  $T$ when they have a clear memory of the outcomes (e.g., see Eq.~\ref{truc}). In \cite{1} it is written that they \textit{apply} quantum mechanics, i.e. (Q), to deduce the outcomes obtained by $W$ and $\overline{W}$. For example if $\overline{F}$ recorded the state $|t\rangle$ then he can infer that $W$ will observe the state $|fail\rangle$ (this is called $\overline{F}^{n:02}$ in \cite{1} and is deduced from Eq.~1 by applying the Born rule $P_{\Psi}(t,ok)=0$). However, this deduction is obtained from an experimental context which is different from the actual experiment described in \cite{1} where  $W$ uses the basis $(ok,fail)$  and  $\overline{W}$  uses the basis $(\overline{ok}, \overline{fail})$. In other words,  $\overline{F}^{n:02}$ `I am certain that W will observe w=fail at time n:31' is contrary to the claim~\cite{1} a counterfactual reasoning which is not justified in quantum mechanics and generally leads to wrong conclusions.  $\overline{F}$  wrong deductions are actually obtained by supposing that the experiment will be the one where  $(t,h)$ and $(ok,fail)$  bases are involved. Moreover, in this alternative experiment  it is not necessary to quantum-eraze the memory of  $\overline{F}$  and the  agent can plays the role of  a genuine observer. For a Bohrian this would be the only  experimental  context in which $\overline{F}$ is a good observer during all the story plot of the experiment. In this perspective the wrong application of quantum mechanics results from the agent $\overline{F}$ belief that he or she completely collapses the wave function given in Eq.~\ref{eq1} and forget the other branches. As we explained earlier this is exactly what usually occurs when the memory is not quantum erased (i.e., with a classical memory).\\ 
\indent Other deductions  shown in Table 3 of \cite{1} are also problematic:
\indent $F^{n:13}$ `I am certain that $\overline{F}$ is certain that $W$ will observe w=fail  at time n:31' and $F^{n:14}$ `I am certain that $W$ will observe w=fail at time n:31 ' are based on (Q) and (C) and are related to $\overline{F}^{n:02}$ wrong application of quantum physics based on a partial knowledge of the experimental context.\\
\indent $F^{n:12}$ `I am certain that $\overline{F}$ knows that $z=+1/2$ at time n:02' is not problematic since it concerns a deduction of  $F$ concerning $\overline{F}$ during a time where his memory  is not quantum erased and where the experimental context is not changed (actually this corresponds to a Rovelian approach~\cite{11}, i.e., to relational quantum mechanics).\\ 
\indent Statement $\overline{W}^{n:22}$ `I am certain that  $F$ knows that $z=+1/2$ at time n:11'  based on the observation $\overline{w}=\overline{ok}$ is problematic and is already discussed in Hardy's paper:  indeed  here $\overline{W}$ applies locally  quantum mechanics  to a context which is not the good one (this leads to the final contradiction) but at least $\overline{W}$ is a good `Bohrian' observer with a memory track of the results.\\
\indent$\overline{W}^{n:23}$ `I am certain that $F$ is certain that $W$  will observe w=fail at time n:31' is based on the previous wrong statements by $F$ so that if $\overline{W}$ knows that  $F$ applies wrongly quantum mechanics but still accept it we can keep this in the Table 3 as  `valid'.  Still,  despite all these wrong applications of quantum mechanics  the local and counterfactual statement $\overline{ok}\rightarrow \uparrow\rightarrow t\rightarrow fail$ is equivalent to $\overline{W}^{n:24}$ 'I am certain that $W$ will observe  w=fail at time $n:31$'. Again this is Hardy's theorem   with the $\textbf{I}+\overline{\textbf{I}}$ inference discussed before.\\
\indent In the run proposed in \cite{1} $\overline{W}$ announces his result (i.e. $\overline{ok}$) and share it with $W$. For this reason  the statement $W^{n:26}$ is unproblematic. Like for the previous deductions $W^{n:27}$  and $W^{n:28}$ are  based on the wrong application of quantum mechanics by previous observers: we can however keep this as `valid' if we know their mistakes (statement $W^{n:28}$ `I am certain that I will observe $w=fail$ at time n:31' is equivalent to $\overline{W}^{n:24}$ and could have been directly written since both $W$ and $\overline{W}$ are on the same observer level for a Bohrian and thus share information in a legitimate way).\\
\indent  All this story plot is apparently very complicated, i.e.,  much more than the initial Hardy's one $\overline{ok}\rightarrow \uparrow\rightarrow t\rightarrow fail$ based only on the locality and non contextuality  assumptions (i.e. the $\textbf{I}+\overline{\textbf{I}}$ inference). As we showed most of the new paradox originates from a wrong application of quantum mechanics taking for granted the role of an observer or agent as a collapsing device (i.e. going to a naive reading of quantum mechanics). These problems disappear together if the observer consider the full wave function needed in the description of the problem and if he or she do not forget the contextual nature of any quantum measurement. The most interesting thing (which was already contained in Deutsch proposal \cite{10} concerns the definition and analysis of the key role of observers as (quantum) memory devices (an idea which is not new and goes back to Everett) and the impact this has on a Bohrian reading of quantum mechanics. For a canonical Bohrian $W$ and $\overline{W}$ are better observers because they can analyze the experiment during all the story. Some Bohrians would however agree that $F$ and
 $\overline{F}$ are allowed to be observers only during the time $T$. Nevertheless, this is not really the official reading of Bohrian mechanics which has pain to define different levels of reality for observation of events. Actually, this way of thinking is more in agreement with the perspective taken by a (neo) Bohrian like a Rovelian (i.e., relational quantum mechanics) which takes more seriously the view taken by several observers (specifically in the context of Einstein's relativity\cite{11,12}) This is also close from the perspective of Qbism as discussed in \cite{1,2}. \\
\indent The previous analysis applies also to the work by C. Brukner \cite{2} where a perfectly entangled pair of spins (i.e., in  the singlet quantum state $|\Psi_{1,2}\rangle=\frac{1}{\sqrt{2}}(|\uparrow_1,\downarrow_2\rangle -|\downarrow_1,\uparrow_2\rangle)$) is observed by two agents like $\overline{F}$ and $F$ (in the bases $(\uparrow_1,\downarrow_1)$ and $(\uparrow_2,\downarrow_2)$ for $\overline{F}$ and $F$ respectively) in a way similar to the one discussed in \cite{1}. Now, using the method proposed in this article and following the proposal of Deutsch \cite{10} the observers can quantum erase their memories and only keep the track that they obtained a definite outcome. We can also imagine that the two agents communicate their results during the game so that we know that they agreed having observed opposed results in the $\pm z$ directions. For example they can write ` We agree having observed a definite outcome. Each of us have obtained an opposite outcome but we can not tell you which one'(we call this message M). Some super observers can now manipulate the two spins still characterized by $|\Psi_{1,2}\rangle$ and can subsequently realize a Bell test violating a inequality. Moreover, if the observers $\overline{F}$ and $F$ believe that they collapsed the wave function (e.g. by ignoring that their memories will be quantum erased) and wrongly apply quantum mechanics they will subsequently deduce that $\overline{W}$ and $W$ should get a result which is \textit{not} violating a Bell inequality in mere contradiction with experimental results by $\overline{W}$ and $W$. Of course, once we know that they apply wrongly quantum laws there is no contradiction. \\
\indent Actually, in \cite{2} this Gedanken experiment was analyzed differently and was proposed in order to exclude the coexistence of `facts' (i.e. measurement outcomes or records) for both the observers and the superobservers (Brukner speaks of observer-independent facts). Using the memory (M) of the definite outcomes we are indeed tempted to fix the spin values along  the $z$ direction (even if we don't remember the precise values) and along any arbitrary directions by using subsequent records obtained by $W$ and $\overline{W}$. Brukner introduces the concept of Boolean algebra and thus claims that the  existence of observer-independent facts implies the existence of joint probabilities $P_{\Psi_{1,2}}(A,A',B,B')$ for different spin observables $A$, $A'$ and $B$, $B'$ recorded by Alice and Bob (the two Wigner friends). This leads to a Bell inequality (i.e., to the Clauser-Horne-Shimony-Holt bound $S=|\langle(A+A')B+(A-A')B'\rangle|\leq 2$) and therefore contradicts quantum mechanics which predicts the Tsirelson bound $S_{\Psi_{1,2}}=2\sqrt{2}$. To be more explicit, the hypothesis of observer-independent facts plays the role of an hidden variable model for testing Bell locality. Here, the  measurement by $\overline{F}$ and $F$ in the $\pm z$ directions leads to the introductions of a local hidden variable 
$\lambda\in [(\uparrow_1,\downarrow_2),(\downarrow_1,\uparrow_2),(\uparrow_1,\uparrow_2),(\downarrow_1,\downarrow_2)]$      
with probability measure $P_{1,2}(\lambda)$ such that $P_{1,2}(\uparrow_1,\downarrow_2)=P_{1,2}(\downarrow_1,\uparrow_2)=1/2$ and $P_{1,2}(\uparrow_1,\uparrow_2)=P_{1,2}(\downarrow_1,\downarrow_2)=0$. Local measurements by Wigner friend Alice involve conditional probabilities defined by the angle $\alpha$ of the Stern and Gerlach apparatus measurement axis with the positive z axis: $P_1(+_\mathbf{a}|\uparrow_1)=cos^2(\alpha/2)=P_1(-_\mathbf{a}|\downarrow_1)$, $P_1(+_\mathbf{a}|\downarrow_1)=sin^2(\alpha/2)=P_1(-_\mathbf{a}|\uparrow_1)$ (which doesn't depend on the state of the second spin $\uparrow_2$, or $\downarrow_2$ which is here omitted). Similar results are obtained on the Bob side recording the second spin state along a direction $\pm\mathbf{b}$. This allows us to define a measurable joint probability $P_{1,2}(x_\mathbf{a},y_\mathbf{b})=\sum_{\lambda}P_1(x_\mathbf{a}|\lambda)P_2(y_\mathbf{b}|\lambda)P_{1,2}(\lambda)$  with  $x,y=\pm 1 $ in agreement with Bell definition of a local and causal hidden variable model. This model satisfies Bell inequality $S\leq 2$ \cite{12b} and therefore contradicts quantum mechanics. Brukner lists four fundamental axioms for deriving this no-go theorem: universal validity of quantum mechanics (i.e. the axiom (Q) of \cite{1}), Bell locality, freedom of choices (i.e., the absence of superdeterminism), and the existence of observer-independent facts (which is similar to (C-S) of \cite{1}) and concludes prudently that, assuming all the other axioms are  satisfied, information taken by the observers and the super observers cannot be taken to coexist~\cite{2}. This view is more balanced than the perspective taken in \cite{1} where the authors conclude on the impossibility of any single-world description of quantum theory (a preliminary version of \cite{1} was untitled `Single-world interpretations of quantum theory cannot be self-consistent' \cite{1b}). Brukner already points out \cite{2} that violation of locality as assumed in Bohmian mechanics  also solve the problem  for his EPR version of the Wigner friend experiment.  We agree with him since  for Bohmian mechanics the existence of observer-independent facts is an ontological postulate.  The strong contextuality and nonlocality of Bohmian mechanics prohibits the possibility to write $P_{1,2}(x_\mathbf{a},y_\mathbf{b})=\sum_{\lambda}P_1(x_\mathbf{a}|\lambda)P_2(y_\mathbf{b}|\lambda)P_{1,2}(\lambda)$ since subsequent measurements made by observers $W$ and $\overline{W}$ (i.e. Alice and Bob) will necessarily include new quantum forces and potentials acting nonlocally on the dynamics of the Qbits. The process (M) on the quantum memory protects the phase coherence of the EPR state and  therefore implies a violation of Bell inequality in agreement with quantum mechanics. We emphasize that Brukner~\cite{2} also included a version of its theorem based on GHZ~\cite{GHZ} nonlocality proof without inequaltiy but with three entangled Qbits. The GHZ theorem brings conclusions similar in philosophy too Hardy's proof. Therefore, the approach advocated by Brukner is less ambiguous than the one provided in \cite{1} using the point of view of several observers applying badly quantum mechanics. The no-go theorem  \cite{2} doesn't contradict Bohmian theory  and we consider it as valid for all nonlocality proofs available in the literature. \\
 \indent A related issue not yet discussed concerns Lorentz invariance which was the main subject of Hardy's article~\cite{3} (see the discussion in \cite{3a}). Indeed, Hardy stressed that while  Bell's theorem  proves that  all realistic interpretations of quantum mechanics must be nonlocal it is natural to ask if they are also non-Lorentz invariant. This makes sense since Bohmian mechanics is non Lorentz invariant and requires a preferred  frame for defining particle trajectories \cite{8} (we will go back to that in  section~\ref{sec:2b}). Hardy~\cite{3} developed a reasoning based on elements of reality a la Einstein  which was criticized in~\cite{3a} because it involves counterfactual deductions neglecting non-locality and contextuality of quantum mechanics. This problem has a certain importance  since in \cite{1,1b,2,2b} one could tempted to use similar deductions with Wigner friends. However, this would lead again to counterfactual contradictions since Wigner friends $\overline{F}$ and $F$  must be treated as quantum memory devices a la Deutsch~\cite{10} and therefore do not escape to the critical analysis and refutations of the early claims made by Hardy~\cite{3a}. Specifically it would be wrong to use deductions made by $\overline{W}$ and $W$ together with Lorentz transformations and moving Lorentz frames (like $\mathcal{F}$ and $\mathcal{F}'$ used in the next subsection) to deduce locally and couterfactually the states of $\overline{F}$ and $F$ in the past (as defined in the common earth reference frame). This would be in conflict with quantum mechanics and Bohmian interpretation.          
\subsection{Bohmian mechanics in a relativistic Universe}
 \label{sec:2b}
 \indent The next important point to be discussed in this article concerns Bohmian trajectories for the proposal \cite{1,1b} (the standard EPR case of \cite{2,2b} leading to Bell's theorem will not be discussed here since one can already find several Bohmian accounts in the literature~\cite{Vigier}). In \cite{1} it is claimed that we can find two camps of Bohmians believing either that  (Q) is unproblematic and (C) is violated or inversely that (C) is unproblematic and (Q) is violated. The authors of \cite{1} do not give too much details~\cite{note} but their conclusion is certainly mistaken: there is only one way to use Bohmian mechanics and there is no ambiguity. In my opinion the main issue is that   as discussed in section \ref{sec:2a} above observers $F$, $\overline{F}$ and $W$ and
 $\overline{W}$ all apply wrongly (i.e. locally and non-contextually) quantum mechanics in many statements of \cite{1} and this leads to Hardy's paradox.  If the observers apply correctly the law of physics then there is no contradiction at all (therefore may be the two camps attribute different meaning to Q and C and this leads to a contradiction).  To illustrate this view I here give the set of all Bohmian trajectories associated with the Qbits involved in Hardy's experiment.  If the various agents know Bohmian quantum mechanics they can predict all the paths followed by the systems during the experiments (here to simplify I admit that the quantum erasing  of the  $F$ and $\overline{F}$ memory process is done so that that I will not have to go back to that point and stick to the historical Hardy's experiment~\cite{3}) up to a subtlety about space time preferred foliations that I will discuss. Also, since these observers know that everything is nonlocal and highly contextual they would obtain the good deductions that a Bohrian  or new-Bohrian (i.e. a Rovelian) should obtain when he/she applies correctly quantum mechanics.\\ 
\begin{figure*}[hbtp]
\includegraphics[width=1\columnwidth]{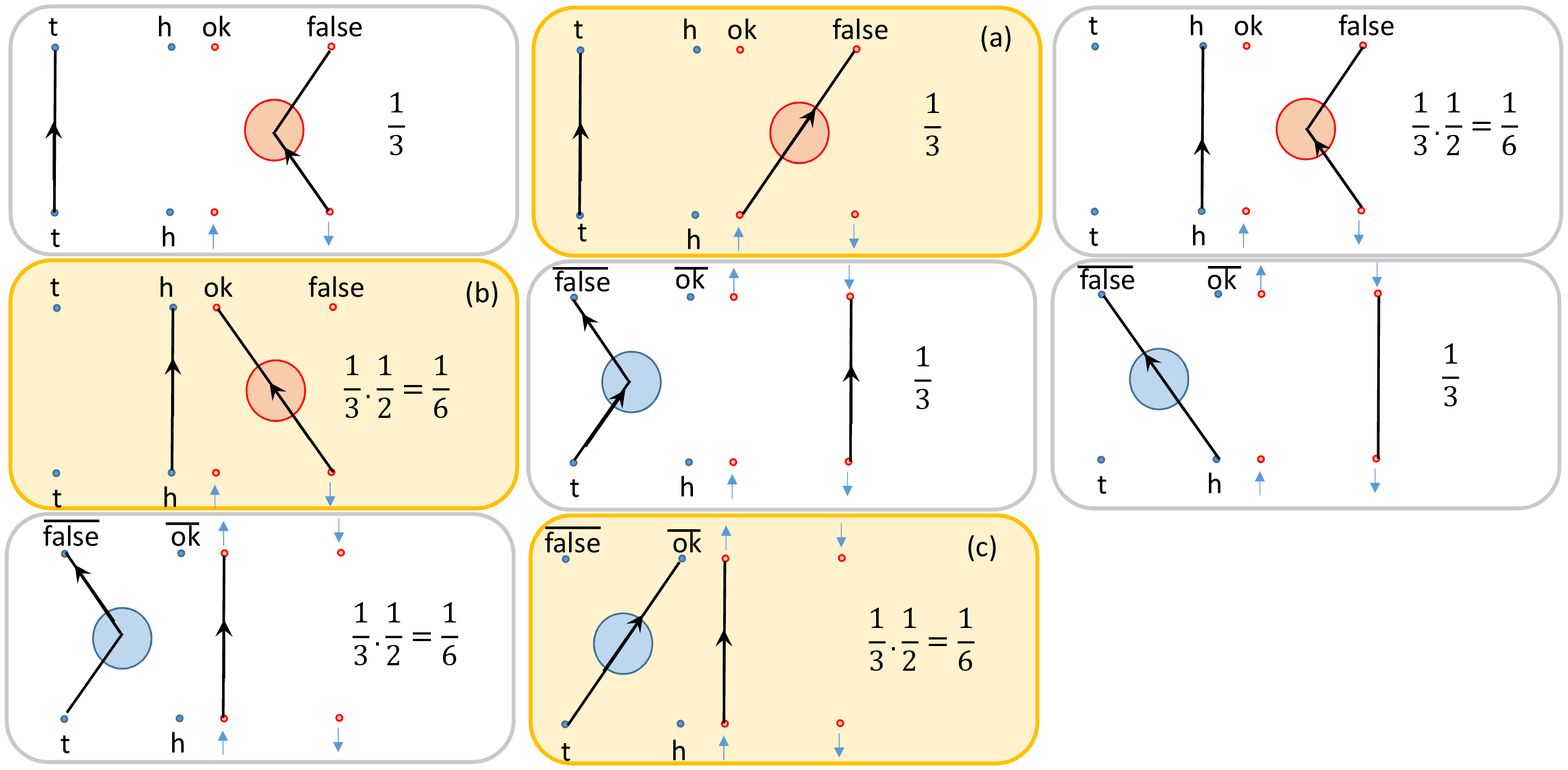}\\
\caption{The panels with a red circle correspond to  experiments where the bases $(t,h)$ and $(ok,fail)$ are actualized (here I used the notation false for fail). The probability associated with Bohmian mechanics are given in each panel (the sum of all probabilities gives one).  For each panel the Bohmian trajectories are represented as a vector (for each panel the time flows vertically from bottom to top and these pictures are more like some kind of Feynman diagrams in a Bohmian world). Same for the panels with a blue circle we have now different experiments where  the     $(\overline{ok},\overline{fail})$ and $(\uparrow,\downarrow)$ bases are used (experiments with red and blue circles are not compatible and correspond to different contexts in a Bohrian sense). A red circle corresponds to some form of linear gate like a beam-splitter allowing us to convert initial states in the basis $(\uparrow,\downarrow)$ to state like in the basis $(ok,fail)$. Similarly the blue circle is a beam splitter mapping states in the basis $(t,h)$ onto the $(\overline{ok},\overline{fail})$ basis. The 3 orange panels are playing a key role in Hardy's paradox~\cite{3,1} and are discussed in the main text.}
\label{fig1}
\end{figure*}
\indent Consider (i.e., Fig.~\ref{fig1}) first the Bohmian trajectories obtained if we use the bases $(t,h)$ and $(ok,fail)$  or if we are using the bases $(\overline{ok},\overline{fail})$ and $(\uparrow,\downarrow)$. These are specific but different experimental contexts described in \cite{3}.  
The orange panel labeled (b) in Fig.~\ref{fig1} corresponds to the situation leading to the inference $ok\rightarrow h$  ($\textbf{I}$) of section \ref{sec:1}. The orange panel (c) in Fig.~\ref{fig1} corresponds to the situation leading to the inference $\overline{ok}\rightarrow \uparrow$  ($\overline{\textbf{I}}$) of section \ref{sec:1}. The local   and noncontextual combination of (b) and (c) leads to $\textbf{I}+\overline{\textbf{I}}$ which is at the core of Hardy's paradox. Equivalently, the panel (c)  can be used with (a) to give the inference $\overline{ok}\rightarrow \uparrow\rightarrow t\rightarrow fail$ discussed in \cite{1} and section \ref{sec:1}. I emphasize that the Bohmian model used here assumes that the initial states $|h\rangle$, and $|t\rangle$ are not spatially overlapping  (same for the states $|\uparrow\rangle$ and $|\downarrow\rangle$ associated with the second Qbit). This hypothesis indeed allows a simple pictorial discussion of Bohmian paths without entering on subtleties about the definition of spins in Bohmian mechanics~\cite{8a,Vigier}. \\ 
\begin{figure*}[hbtp]
\includegraphics[width=1\columnwidth]{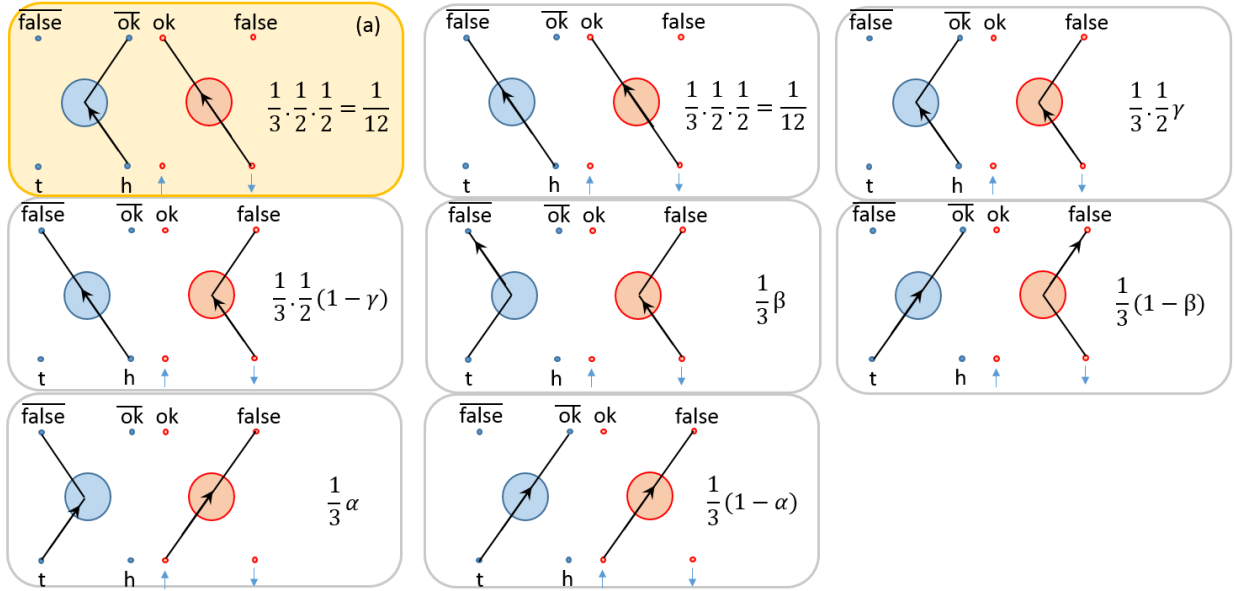}\\
\caption{The different panels show Bohmian paths graphically represented when the agents $W$ and $\overline{W}$ use the bases  $(ok,fail)$ and   $(\overline{ok},\overline{fail})$ in their measurement. The paths are calculated by using the preferred foliation  $\mathcal{F}$ discussed in the main text. The symbols and notations are otherwise the same as in Fig.~\ref{fig1}. The orange panel labeled (a) corresponds to the outcomes  $ok,\overline{ok}$ of Eq.~\ref{eq2} playing a central role in Hardy's paradox (compare with Fig.~\ref{fig3}). The probabilities are evaluated using conservation law  and causality seen from the point of view of $\mathcal{F}$ (the coefficients $\alpha,\beta,\gamma\in[0,1]$ obey to the sum rule $7=4(\alpha+\beta)-2\gamma$).   }
\label{fig2}
\end{figure*}
\indent Now, if we want to draw Bohmian paths associated with the experiments where the observers $W$ and $\overline{W}$ use the bases  $(ok,fail)$ and   $(\overline{ok},\overline{fail})$ we have to be more prudent. Indeed, Bohmian mechanics is a nonlocal theory and for an experiment like the one of Hardy \cite{1,3} we have to care about the Lorentz frame used to calculate the Bohmian paths. More precisely, 
\begin{figure*}[hbtp]
\includegraphics[width=1\columnwidth]{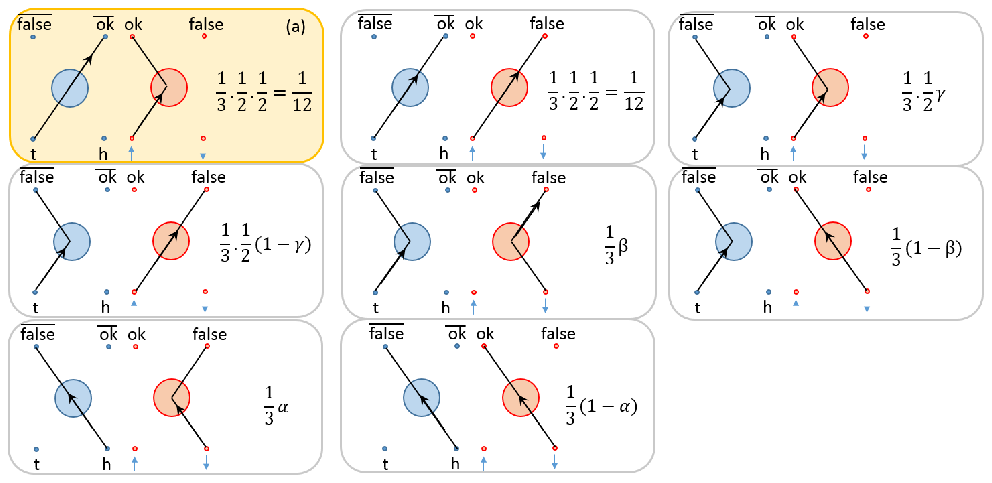}\\
\caption{Like for Fig.~\ref{fig2} the different panels show Bohmian paths graphically represented when the agents $W$ and $\overline{W}$ use the bases  $(ok,fail)$ and   $(\overline{ok},\overline{fail})$ in their measurement. The paths are now calculated by using the preferred foliation  $\mathcal{F}'\neq\mathcal{F} $ discussed in the main text. The symbols and notations are otherwise the same as in Fig.~\ref{fig1}. The orange panel labeled (a) corresponds to the outcomes  $ok,\overline{ok}$ of Eq.~\ref{eq2} playing a central role in Hardy's paradox (compare with Fig.~\ref{fig2}). }
\label{fig3}
\end{figure*}
Bohmian mechanics require a preferred Lorentz frame in which a time ordering  will be defined for calculating the paths~\cite{8a,9,13}. But since the  events associated  with local labs $L$ (i.e. $W$)   and $\overline{L}$ (i.e. $\overline{W}$) are space-like  separated  the order of events could be different in different Lorentz frame and the  Bohmian paths could be different (i.e. the path are foliation dependent). This strongly impacts the notion of probability of presence and equivariance in Bohmian mechanics~\cite{9,13}.\\
\indent In Fig.~\ref{fig2} we show what will happen if we suppose that the preferred frame used to calculate the path is such that the agent $W$ detects the particle (spin) in the basis $(ok,fail)$ before the entangled particle (coin) cross the other beam splitter (in blue). In other words in this reference frame $\mathcal{F}$ the coin is still in a $t$ or $h$ state  while the spin is detected by $W$. This allows simple graphical representations of the possible Bohmian paths  $X_{\mathcal{F}}(t)$ for the two Qbits (shown in Fig.~\ref{fig2} in the laboratory reference frame where $W$ and $\overline{W}$ are simultaneous). From all the possible trajectory sets  the orange panel labeled (a)  plays a key role in Hardy's experiment since it corresponds to the outcome $ok,\overline{ok}$ with the probability  $P_\Psi(ok,\overline{ok})=\frac{1}{12}$ of Eq.~\ref{eq2}. Importantly, this is the only panel of Fig.~\ref{fig2} where such an outcome occurs. It is associated with  a coin starting in the $h$ state while the spin is in the $\downarrow$ state.  This contradicts the $\textbf{I}+\overline{\textbf{I}}$ inference  which prohibits such a possibility. Again nonlocality is at play here and trajectories are strongly modified by the change of contexts even in space-like  separated regions of the Universe.\\   
\indent That's not all. In a Bohmian world we can also use the alternative preferred foliation $\mathcal{F}'$~\cite{remark} where the agent $\overline{W}$ detects the coin before the spin even reached the red beam splitter. In this alternative foliation  $\mathcal{F}'$ (see Fig.~\ref{fig3}) we can also graphically represents (in the laboratory reference frame where $W$ and $\overline{W}$ are simultaneous) the possible Bohmian paths associated  with the same experiment in the $(ok,fail)$ and $(\overline{ok},\overline{fail})$ bases. Remarkably the paths $X_{\mathcal{F}'}(t)$ and $X_{\mathcal{F}}(t)$ are different and associated with different particle distributions (i.e., Born's rule is foliation dependent at the hidden variable level).  The most important feature is again the orange panel labeled (a) in  Fig.~\ref{fig3} which shows trajectories ending in the $ok,\overline{ok}$ with the probability  $P_\Psi(ok,\overline{ok})=\frac{1}{12}$ of Eq.~\ref{eq2}. This is again the only panel of Fig.~\ref{fig3} associated with such an outcome and we can see that these paths starting in the  $t,\uparrow$  are radically different from those obtained in the orange panel (a) of Fig.~\ref{fig2}! This confirms that different foliation in this nonlocal Bohmian theory implies in general different trajectories for entangled states~\cite{9,13,14}.\\
\indent The involvement of foliations in relativistic Bohmian dynamics \cite{9,13} is central in the understanding of such a theory. For the present discussion  it plays a key role for agents knowing Bohmian mechanics and trying to calculate the paths they follow.  Imagine that the agents $F$ and $\overline{F}$ know that they are in the $t,\uparrow$ state. As we know from statement $F^{n:12}$ this is completely allowed. If the agents know that the wave function guiding their paths is given by Eq.~\ref{eq1} they  can predict the outcomes of the experiments with the different probabilities (see Figs.~\ref{fig2},\ref{fig3}). However, in order to define precisely the dynamics they also need to know which foliation plays a preferred role in the dynamics, i.e., they need to know which Lorentz frame is a `subquantum Aether' a la Bohm-Vigier~\cite{8a}. If they don't know this frame they  will not be able to define univocally the paths.                                                               
\section{Conclusion}
\indent To summarize we discussed the proposal \cite{1} based on Hardy's paradox  and showed that the analysis in term of Wigner friends and agents doesn't lead to new paradox not already contained in the previous works about nonlocality and contextuality of quantum mechanics. We showed that many paradoxical statements in \cite{1,1b} are actually deduced by agents who are badly applying quantum mechanics. Those agents, ignoring the actual evolution of the wave function and experiments, forget that they are themselves genuine quantum systems with quantum memories and this can induce apparent contradictions.
Of course, the realization of genuine quantum memories a la Deutsch \cite{10} is a difficult technical issue but nothing prohibits us to develop `baby-Wigner' friends more in the direction of quantum which-path or quantum-eraser experiments involving only photons or like the Schrodinger kittens using Rydberg atoms and developed by Serge Haroche group. Moreover, any Bohrian or Rovelian agent applying correctly quantum mechanics can infer that `by applying correctly quantum rules, i.e., by taking into account the full wave function $|\Psi(t)\rangle$ of the system I can predict unambiguously the outcome probabilities for experiments made by other agents'. Furthermore the same agent can deduce fairly  that `if I have a quantum memory which will be erased during the protocol I will not be able to violate Bohr's complementarity or the uncertainty principle'. In other words, the mere fact that Wigner friends $F$ and $\overline{F}$ remember that they participated to the experiment \cite{1} or \cite{2} but can not tell precisely which outcomes occurred  protects the self consistency of Bohr complementarity which is all about information available to an observer not about ontology of the hidden world. To paraphrase Bell: `complementary is safe FAPP' (i.e., for all practical purposes).\\
\indent We also showed that there is no contradiction in \cite{1,2} forbidding us to apply a `single-world' interpretation of quantum mechanics such as Bohmian mechanics~\cite{Sudbery} to any experiment involving one or several Wigner and Wigner friends. A Bohmian agent can fairly states: `knowing the quantum state $|\Psi(t)\rangle$ of the system I will be able not only to predict the probability outcomes (like for a Bohrian agent) but I will also be able to deduce the complete dynamics and trajectories of Qbits and agents involved in the process'. Furthermore, he or she could add: ` While my quantum memory can be erased during the process I will forgot which state I actually had and this will be done in order to protect complementarity and the Heisenberg principle for any other observers'. Therefore, at the end a Bohmian can be a Bohrian/Rovelian FAPP but the reverse is certainly not true. In my option having a clear deterministic vision of a quantum dynamics  a la de Broglie Bohm helps for giving a clean foundation to the orthodox interpretation (this was already claimed by Bohm~\cite{8a}). Moreover, this will be true even if at the end we accept that a part of the ontology is hidden or protected from our intervention as agents or observers.    
\acknowledgments
I thank Renato Renner for some illuminating comments and Dustin Lazarovici for sharing with me his recent work~\cite{dustin} where conclusions close to those contained in this article were independently obtained.         
   


\begin{references}
\bibitem{comment} The states $|\overline{ok}\rangle=\frac{|h\rangle-|t\rangle}{\sqrt{2}}$ etc... are actually time dependent and we can alternatively use the definition $|\overline{ok}(t)\rangle=U_t[\frac{|h\rangle-|t\rangle}{\sqrt{2}}]$ where $U_t$ is the unitary time-evolution.
\bibitem{5} We have $P_\Psi(t,fail)=2/3$, $P_\Psi(h,ok)=P_\Psi(h,fail)=1/6$. Similarly $P_\Psi(\overline{fail},\downarrow)=2/3$, $P_\Psi(\overline{ok},\uparrow)=P_\Psi(\overline{fail},\uparrow)=1/6$.
\bibitem{12} Still for Rovelians in order to give a meaning to other observer perspectives means  that they, at least, all agree on the existence of a common hidden reality that they are all watching from a different side.  This hidden reality is exactly what a Bohmian wants to describe.
\bibitem{12b} We even see simpler contradictions with quantum mechanics since we have  $P_{1,2}(+_\mathbf{a},-_\mathbf{a})=\frac{1}{2}cos^4(\alpha/2)$, $P_{1,2}(+_\mathbf{a},+_\mathbf{a})=cos^2(\alpha/2)sin^2(\alpha/2)$ to be compared with quantum mechanics predictions $P_{\Psi_{1,2}}(+_\mathbf{a},-_\mathbf{a})=\frac{1}{2}$ and $P_{\Psi_{1,2}}(+_\mathbf{a},+_\mathbf{a})=0$.
\bibitem{note} In \cite{1} it is written: `According to the common understanding, Bohmian mechanics is a "theory of the universe" rather than a theory about subsystems. This means that agents who apply the theory must in principle always take an outside perspective on the entire universe, describing themselves as part of it. This outside perspective is identical for all agents, which ensures consistency and hence the validity of Assumption (C). However, because (S) is satisfied, too, it follows from Theorem 1 that (Q) must be violated'. They also wrote (to define a second Bohmian camp): `Indeed, if the agents applied the Bohmian equations of motion directly to the relevant systems around them, rather than to the universe as a whole, their reasoning would be the same as the one prescribed by (Q). But since Bohmian mechanics also satisfies (S), this would, by virtue of Theorem 1, imply a violation of (C), i.e., the agents’ conclusions would contradict each other'.
\bibitem{remark} The definition of $\mathcal{F}$, and $\mathcal{F}'$ require Lorentz transformation such that
\bibitem{14} We considered two families of foliations $\mathcal{F}$ and $\mathcal{F}'$ which differ  by the time ordering of some events in their respective preferred Lorentz frames. There is also a degenerate  foliation  where $\mathcal{F}=\mathcal{F}'$ i.e. the  common laboratory reference frame.  In this foliation the trajectory topology are the one appearing in both Fig.~\ref{fig2} and Fig.~\ref{fig3}. For example  the trajectories in the orange panel of Fig.~\ref{fig2} and Fig.~\ref{fig3} are present with a probability weight $1/24$ (the sum leads to 1/12).
\end{references}
\end{document}